\documentclass[a4paper,twoside,10pt]{letter}
\usepackage{graphicx,saj,multicol,subeqnarray}

\def\papertype{}

\def\udc{...}
\begin{document}
\baselineskip=3.1truemm
\columnsep=.5truecm
\newenvironment{lefteqnarray}{\arraycolsep=0pt\begin{eqnarray}}
{\end{eqnarray}\protect\aftergroup\ignorespaces}
\newenvironment{lefteqnarray*}{\arraycolsep=0pt\begin{eqnarray*}}
{\end{eqnarray*}\protect\aftergroup\ignorespaces}
\newenvironment{leftsubeqnarray}{\arraycolsep=0pt\begin{subeqnarray}}
{\end{subeqnarray}\protect\aftergroup\ignorespaces}
%


\markboth{\eightrm OPTICAL OBSERVATIONS OF M81 GALAXY GROUP}
{\eightrm B. ARBUTINA et al.}

{\ }

\publ

\type

{\ }


\title{OPTICAL OBSERVATIONS OF M81 GALAXY GROUP IN
NARROW BAND [SII] AND H$\alpha$ FILTERS: HOLMBERG IX}


\authors{B. Arbutina$^{1}$, D. Ili\'{c}$^{1}$, K. Stavrev$^{2}$, D. Uro{\v s}evi{\' c}$^{1}$ B. Vukoti\'{c}$^3$ and D. Oni\' c$^1$}

\vskip3mm


\address{$^1$Department of Astronomy, Faculty of Mathematics,
University of Belgrade,\break Studentski trg 16, 11000 Belgrade,
Serbia}

\Email{arbo}{matf.bg.ac.rs}

\address{$^2$Institute of Astronomy, Bulgarian Academy of Sciences, 72 Tsarigradsko Shosse Blvd., \break BG-1784 Sofia, Bulgaria}

\address{$^3$Astronomical Observatory, Volgina 7, 11160 Belgrade,
Serbia}



\dates{Jun 15, 2009}{December ?, 2009}


\summary{We present observations of the nearby tidal dwarf galaxy
Holmberg IX in M81 galaxy group in narrow band [SII] and H$\alpha$
filters, carried out in March and November 2008 with the 2m RCC
telescope at NAO Rozhen, Bulgaria. {Our search for resident
supernova remnants (as identified as sources with enhanced [SII]
emission relative to their H$\alpha$ emission) in this galaxy has
yielded no sources of this class, besides M\&H 10-11 or HoIX X-1.}
Nevertheless we found a number of objects with significant
H$\alpha$ emission that probably represent uncatalogued HII
regions. }


\keywords{ISM: supernova remnants -- HII regions -- Methods:
observational -- Techniques: photometric -- Galaxies: ISM --
Galaxies: individual: Holmberg IX}

\begin{multicols}{2}
{


\section{1. INTRODUCTION}

M81 galaxy group is the nearest interacting group of galaxies
whose main members are M81, M82 and NGC 3077. Yun et al.~(1994)
found prominent HI structures surrounding these galaxies with
large HI complexes and tidal bridges, that are probably a result
of the galaxy encounters. It is possible that starburst activity
i.e. enhanced star formation of M82 was triggered in a close
encounter with M81, which as a consequence have a high supernova
rate (see e.g. Arbutina et al. 2007, Huang et al. 1994). {The
third member in the group, NGC 3077
 also shows evidence of enhanced star formation, and
consequently a higher supernova rate and presence of SNRs. This
was partially confirmed by recent radio observations (see
Rosa-Gonzales 2005). } There is also a number of optical
candidates for supernova remnants (SNRs) detected in M81 (Matonick
\& Fesen 1997). The SNR candidates in optical are usually
identified through enhanced [SII] line emission ([SII]/H$\alpha >
0.4$; see e.g. Matonick \& Fesen 1997, Blair \& Long 2004). The
aim of our optical observations was to try to detect new SNR
candidates and HII regions in a small "satellite galaxy" of M81 --
Holmberg IX (UGC 5336, MCG+12-10-012, LEDA 28757, see Table 1).
Holmberg IX and Arp's loop can be seen as two dark knots in the HI
image of the M81 triplet of Yun et al. (1994). Dwarf irregular
galaxy Holmberg IX could be the youngest nearby tidal dwarf
galaxy, perhaps formed during the last close passage of M82 around
M81 (Sabbi et al. 2008).

}
\end{multicols}

\vfill\eject


\centerline{\includegraphics[bb=5 0 410 320,
keepaspectratio,width=16cm]{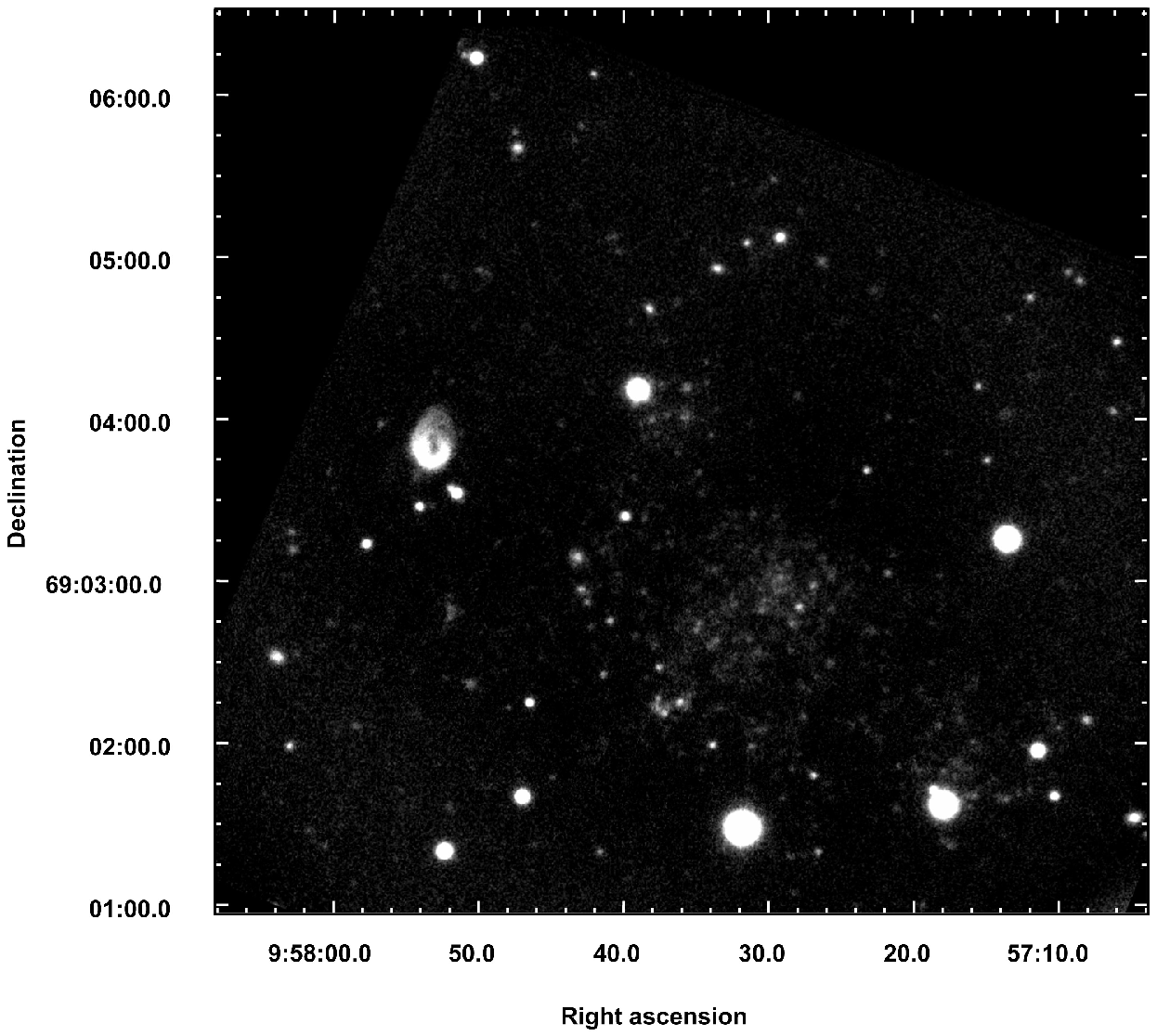}}


\vspace{3mm}
\begin{minipage}{14.5cm}
\centering { \figurecaption{1.}{{The [SII]} with continuum image
(sky subtracted).} }
\end{minipage}
\vspace{9mm}

\noindent
\begin{minipage}{\textwidth} \centering
 {\bf Table 1.} Data for Holmberg IX (MCG+12-10-012) taken from
SIMBAD$^\dag$. \vskip5mm
\centerline{\begin{tabular}{@{\extracolsep{-2.0mm}}c c c c c c c c
@{}} \hline
 Right ascension  & Declination & Redshift & Velocity & Distance$^\ddag$ & Angular size  & Magnitude & Morphological \\
  $\alpha _{\mathrm{J2000}} $ & $\delta _{\mathrm{J2000}}$ & $z$ & $v$ [km s$^{-1}$] & $d$ [Mpc] & [ $'$ ]& &  type \\
\hline \hline
 09 57 32.1 & +69 02 46  & 0.000213 & 64 & 3.7 &
 $2.6 \times 2.2$ &  16.5 (B) & dI \\
\hline
\end{tabular}}
\vskip2mm \hskip-50mm $^\dag${\footnotesize
\texttt{http://simbad.u-strasbg.fr/simbad/}}\ \ \
$^\ddag${\footnotesize Karachentsev \& Kashibadze (2006)}
\vspace{2mm}
\end{minipage}

 \vskip.95cm

\begin{multicols}{2}
{

{The adopted distance to Holmberg IX is $d=3.7$ Mpc, which is the
distance derived from cepheids distance to M81 and known
membership in the M81 group (Karachentsev et al. 2004,
Karachentsev \& Kashibadze 2006).} The location of Holmberg IX,
its high gas content, and its youthful stellar population, made
from this galaxy the primary target of our search.

 \vskip.5cm

\centerline{{\bf Table 2.} Characteristics of the narrow band
filters.} \vskip5mm \centerline{\begin{tabular}{c | c c c }
Filter & $\lambda _o \ \mathrm{[\AA\mathrm ]}$ & FWHM $\mathrm{[\AA\mathrm ]}$ & $\tau _\mathrm{max}$ [\%] \\
\hline \hline
 [SII] & 6719 & 33 & 83.3 \\
 H$\alpha$ & 6572 & 32 & 86.7 \\
 Red cont. & 6416 & 26 & 58.0 \\
\hline
\end{tabular}}

\vskip.5cm

}
\end{multicols}

\vfill\eject

\centerline{\includegraphics[bb=5 0 410 320,
keepaspectratio,width=16cm]{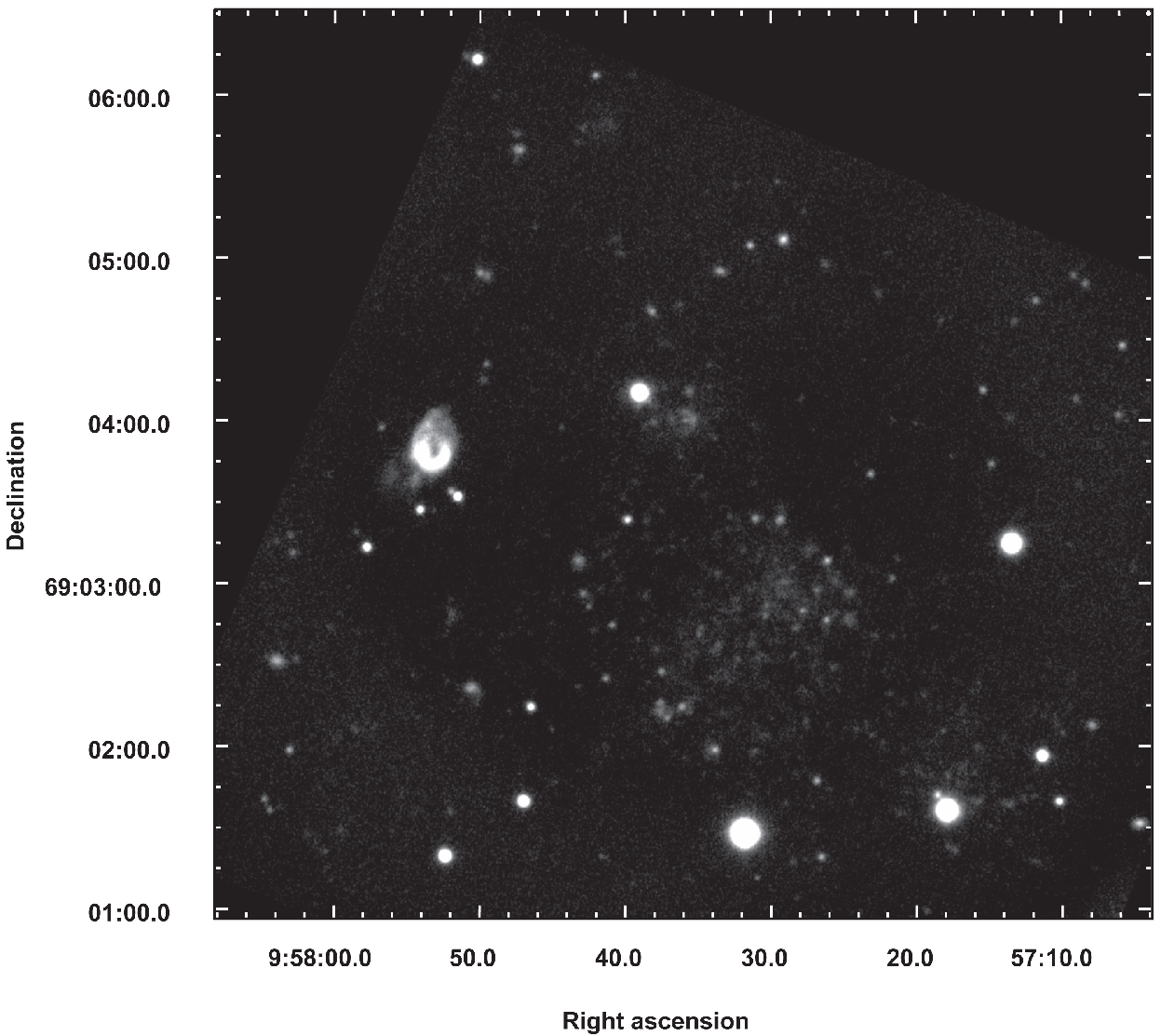}}


\vspace{4mm}
\begin{minipage}{14.5cm}
\centering
 \figurecaption{2.}{The H$\alpha$ with continuum image (sky
subtracted).} \vspace{5mm}
\end{minipage}

\vspace{5mm}

\noindent

 \begin{minipage}{14.5cm}
 \centering
  {{\bf Table 3.} The observations log. }
  \vskip 5mm
  \centerline{
   \begin{tabular}{@{\extracolsep{+5.0mm}}l | ccc | ccc@{}}
   \hline
             & \multicolumn{6}{c}{Integration time [s]}  \\
   Object/SS   & \multicolumn{3}{c}{2008 March 3} & \multicolumn{3}{c}{2008 November 30}  \\
             & {\footnotesize Cont.}&  {\footnotesize H$\alpha$} & {\footnotesize [SII]} & {\footnotesize Cont.}&  {\footnotesize H$\alpha$} & {\footnotesize [SII]}   \\
 \hline \hline
                     & 1800 & 1200 & 1200 & 1800 & 1200 & 1200 \\
 Holmberg IX         & 1800 & 1800 & 1800 & 1800 & 1200 & 1200 \\
                     & 2400 & 1800 & 1800 & 1800 & 1200 & 1200 \\
 \hline
                     & 180  & 120  & 120 & -- & -- & -- \\
 Feige 34            & 240  & 120  & 120 & -- & -- & -- \\
                     & 180  & 120  & 120 & -- & -- & -- \\
 \hline
                     & --  & --  & -- & 300 & 120 & 120 \\
 G191-B2B            & --  & --  & -- & 300 & 120 & 120 \\
                     & --  & --  & -- & 300 & 120 & 120 \\
 \hline
\end{tabular}
}
\end{minipage}

\vfill\eject


\centerline{\includegraphics[bb=5 0 410 320,
keepaspectratio,width=16cm]{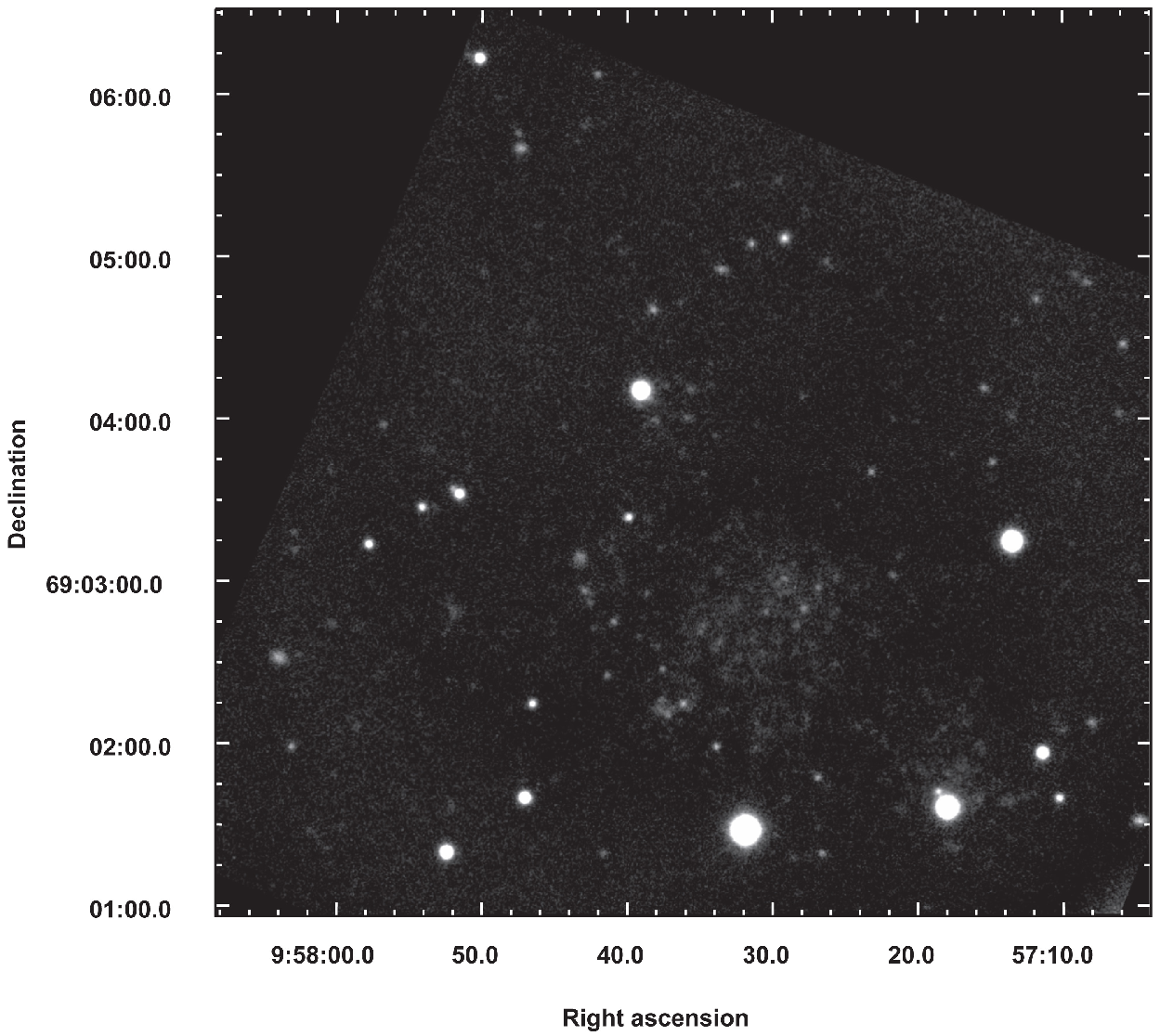}}


\vspace{1mm}
\begin{minipage}{14.5cm}
\centering
 \figurecaption{3.}{The red continuum image (sky subtracted).}
\end{minipage}
\vspace{3mm}

\begin{multicols}{2}

\section{2. OBSERVATIONS AND DATA REDUCTION}

The observation were carried  out in March and November 2008 with
the 2 m Ritchey-Chr\'{e}tien-Coud\'{e} (RCC) telescope at the
National Astronomical Observatory (NAO) Rozhen, Bulgaria ($\varphi
= 41^\circ 41'\ 35'' ,\ \lambda = 24^\circ 44'\ 30'' ,\ h = 1759$
m). In the RC focus of the telescope, the equivalent focal length
is 16 m and the field-of-view is one square degree with a scale
12\uu 89/mm. The telescope is equipped with VersArray: 1300B CCD
camera with 1340$\times$1300 px array, with plate scale of 0\uu
257732/px (pixel size is 20 $\mu$m), giving field of view
$5'45''\times 5'35''$.

We used the narrow-band filters for [SII], H$\alpha$ and red
continuum. We took sets of three images through each filter, each
night, with exposure time ranging from 1200--2400 s. Typical
seeing was 1\uu 5 -- $3''$. Standard stars images (Feige 34 and
G191-B2B), bias frames and sky flat-fields were also taken.
Filters characteristics and details of the observations are given
in Tables 2 and 3. The H$\alpha$ image ($\lambda 6563$) is
contaminated with some [NII] emission ($\lambda 6583$), so in
principle the "H$\alpha$" is actually H$\alpha$+[NII] image. The
[SII] filter should collect most of the emission from both [SII]
$\lambda 6716$ and $\lambda 6731$ lines.

Data reduction was performed by using IRIS\footnote{Available from
{\footnotesize \texttt{http://www.astrosurf.com/buil/}}} (an
astronomical images processing software developed by Christian
Buil). The data were bias subtracted and flat-fielded using the
standard methods. Three images in each set are  combined using
procedures \texttt{NGAIN3} and \texttt{COMPOSIT}, and then
sky-substructed (\texttt{SUBSKY}). Commands {\texttt{MAX, MIN,
EDGE}} were used for cosmetic corrections (bad pixels, cosmic rays
removal). Since the images were taken with different exposures,
depending on the filters, we scaled all the images normalizing
them to the flux of the stars in the field. Images taken on two
nights are then combined to increase the counts, rotated, aligned
and distortion corrected by using the procedure
\texttt{COREGISTER}. Images obtained after all these corrections,
are given in Figs. 1--3.

\end{multicols}


\centerline{\includegraphics[bb=0 0 1340 1300,
keepaspectratio,width=16cm]{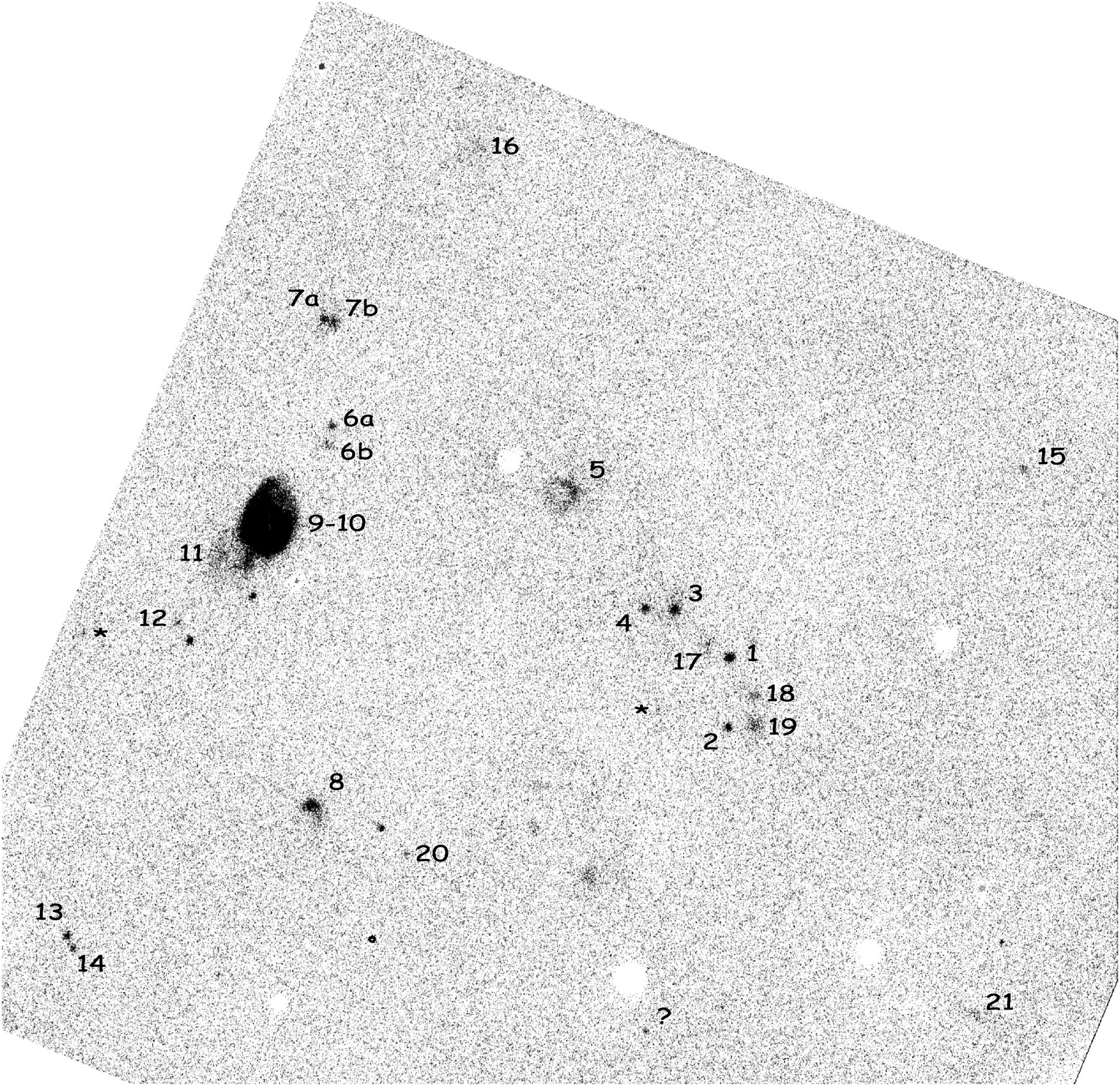}}


\vspace{4mm}
\begin{minipage}{14.5cm}
 \figurecaption{4.}{The continuum-subtracted H$\alpha$ image. Dark features that are marked are sources. Additional dark features are stars not subtracted well (compare Figs. 2 and 3).}
\end{minipage}

 \vspace{3mm}

\begin{multicols}{2}

The H$\alpha$ and [SII] images are then continuum-subtracted and
final images once again background subtracted to obtain background
as flat as possible. The continuum-subtracted H$\alpha$ image is
given in Fig. 4.

Sources in the continuum-subtracted H$\alpha$ image are extracted
by smoothing the image and then drawing 1$\sigma$ contours from
the median value. Relative fluxes (total counts) are then
calculated using IRIS photometric tools. Finally, an astrometric
reduction of the H$\alpha$ image was performed by using U.S. Naval
Observatory's USNO-A2.0 astrometric catalogue (Monet et al. 1998).

\section{3. ANALYSIS  AND  RESULTS}

The continuum-subtracted [SII] image did not show any new object
with an enhanced [SII] emission, besides M\&H 10-11 or HoIX X-1
(Miller \& Hodge 1994, Miller 1995, Gris\'{e} et al.~2006), a
strong optical line source and ultraluminous X-ray source (a
possible hypernova remnant or super-shell), and thereby is
omitted. As for the continuum-subtracted H$\alpha$ image, we
detected 21 sources -- probable HII regions (see Fig. 4;
additional dark features are stars not subtracted well).

\vskip1.5cm

\centerline{\includegraphics[bb=15 20 270 230,
keepaspectratio,width=8cm]{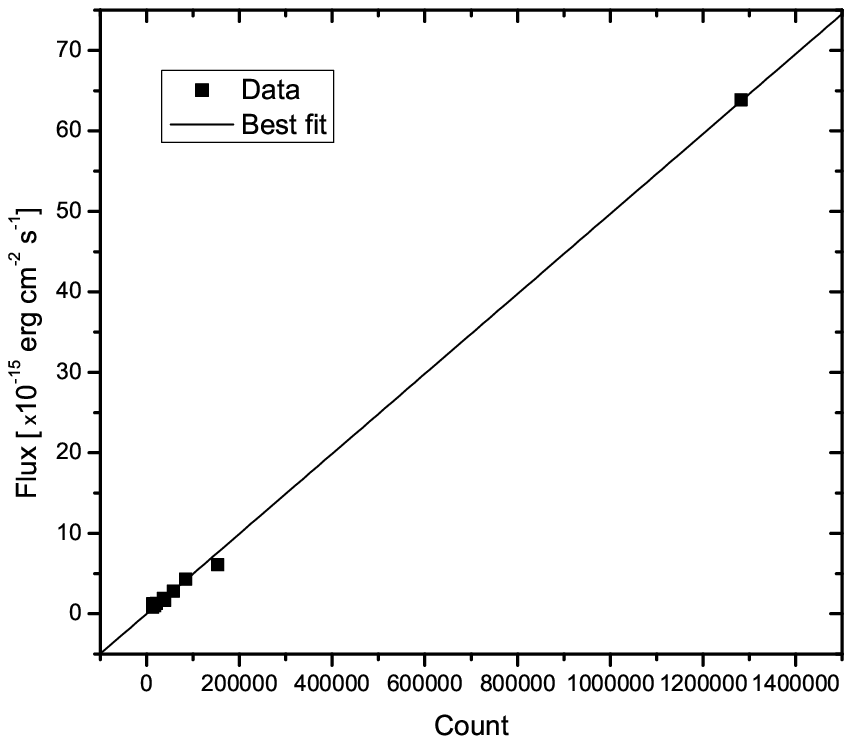}}

\vspace{1mm} \noindent
\begin{minipage}{8.0cm}
 \figurecaption{5.}{Flux -- Count ($F-N$) relation used for absolute calibration.
 Slope of the relation is calibration coefficient: $c=(4.97 \pm 0.05) \times 10^{-5}$.}
\end{minipage}
\vspace{6mm}

\noindent \begin{minipage}{8.0cm}
 \centerline{{\bf Table 4.} Sources used for absolute
calibration.} \vskip5mm
\centerline{\begin{tabular}{@{\extracolsep{+2.5mm}}l |  c c c c c
@{}}
      &        & \multicolumn{4}{c}{H$\alpha$ Flux }  \\
Source & Count & \multicolumn{4}{c}{\footnotesize [$\times 10^{-15}$ erg cm$^{-2}$ s$^{-1}$] }  \\
          &  {\footnotesize this paper}  &  \multicolumn{4}{c}{\footnotesize from M\&H }  \\
\hline \hline
M\&H 1   &   20721   && &1.30 &   \\
M\&H 2   &   12884   && &1.27 &    \\
M\&H 3   &   38559   && &1.65 &     \\
M\&H 4   &   17229   && &0.99 &     \\
M\&H 5   &   83939   && &4.32 &     \\
M\&H 6   &   12807   && &0.83 &      \\
M\&H 7   &   36059   && &1.92 &       \\
M\&H 8   &   57565   && &2.83 &       \\
M\&H 9-10    &   1282274 && &63.87 &     \\
M\&H 11  &   153357  && &6.10 &    \\
\hline
\end{tabular}}
\end{minipage}
 \vspace{6mm}


Eleven sources (1--11) were previously identified by Miller \&
Hodge (1994) and one by Boone et al.~(2005) (source
12).\footnote{See Aladin Sky Atlas: {\footnotesize
\texttt{http://aladin.u-strasbg.fr/}}.} Thus, we found nine new
sources (13--21).

\end{multicols}

\vskip0.9cm

\noindent
\begin{minipage}{\textwidth} \centering
 {\bf Table 5.} HII regions in Holmberg IX.
 \vskip5mm \centerline{\begin{tabular}{@{\extracolsep{3.0mm}}l |
c c c c  @{}}
Source & Right ascension  & Declination &  H$\alpha$ Flux  & Comment\\
 &  $\alpha _{\mathrm{J2000}} $ & $\delta _{\mathrm{J2000}}$ &  $F_{\mathrm{H}\alpha}$ [$\times 10^{-15}$ erg cm$^{-2}$ s$^{-1}$] & \\
\hline \hline
Aea 1   &   09 57 26.1  &   +69 03 09 &         1.03 & M\&H 1 \\
Aea 2   &   09 57 26.2  &   +69 02 47 &         0.64 & M\&H 2 \\
Aea 3   &   09 57 29.4  &   +69 03 24 &         1.92 & M\&H 3 \\
Aea 4   &   09 57 31.1  &   +69 03 24 &         0.86 & M\&H 4 \\
Aea 5   &   09 57 35.6  &   +69 04 01 &         4.17 & M\&H 5 \\
Aea 6A  &   09 57 49.5  &   +69 04 21 &         0.38 & M\&H 6 \\
Aea 6B  &   09 67 49.8  &   +69 04 16 &         0.26 & M\&H 6  \\
Aea 7A  &   09 57 49.9  &   +69 04 55 &         1.03 & M\&H 7 \\
Aea 7B  &   09 57 49.5  &   +69 04 54 &         0.76 & M\&H 7 \\
Aea 8   &   09 57 50.5  &   +69 02 22 &         2.86 & M\&H 8 \\
Aea 9-10$^a$    &   09 57 53.1  &   +69 03 49 &    63.71 & M\&H 9-10 \\
Aea 11  &   09 57 55.2  &   +69 03 40 &         7.62 & M\&H 11 \\
Aea 12  &   09 57 58.5  &   +69 03 19 &         0.24 & $-$\\
Aea 13  &   09 58 04.7  &   +69 01 41 &         0.56 & $-$\\
Aea 14  &   09 58 04.4  &   +69 01 37 &         0.36 & $-$\\
Aea 15  &   09 57 09.0  &   +69 04 08 &         0.33 & $-$\\
Aea 16  &   09 57 41.0  &   +69 05 50 &         0.70 & $-$\\
Aea 17  &   09 57 27.4  &   +69 03 13 &         0.43 & $-$\\
Aea 18  &   09 57 24.6  &   +69 02 57 &         0.55 & $-$\\
Aea 19  &   09 54 24.6  &   +69 02 47 &         1.13 & $-$\\
Aea 20  &   09 57 45.0  &   +69 02 07 &         0.13 & $-$\\
Aea 21  &   09 57 11.8  &   +69 01 18 &         0.29 & $-$\\
\hline
\end{tabular}}
\vskip2mm \hskip-80mm
 $^a${\footnotesize Possible hypernova remnant or
super-shell.}
\end{minipage}

\clearpage

\begin{multicols}{2}
{

Additionally, we resolved sources M\&H 6 and M\&H 7, which we see
as two pairs. There are, possibly, two more smaller sources,
marked with asterisk (*) in Fig.~4, at $\alpha _{\mathrm{J2000}}$
= $09^h 57^m 30^s.3 $, $\delta _{\mathrm{J2000}}$ = $+69^\circ 02'
52''$ and $\alpha _{\mathrm{J2000}}$ = $09^h 58^m 04^s.0 $,
$\delta _{\mathrm{J2000}}$ = $+69^\circ 03' 16'' $, for which we
haven't measured fluxes. Question mark (?) in Fig. 4 marks the
position of a source of unknown origin ($\alpha _{\mathrm{J2000}}$
= $09^h 57^m 30^s.9 $, $\delta _{\mathrm{J2000}}$ = $+69^\circ 01'
12'' $) which we saw in March \ 2008 H$\alpha$ images, but not in
November 2008.

The absolute flux calibration of the continuum subtracted
H$\alpha$ image was performed by using the fluxes of sources
identified both by us and Miller \& Hodge (1994) (See Fig. 5 and
Table 4). {The listed fluxes for sources M\&H 6 and M\&H 7
correspond to the sum of the fluxes for the two sub-regions (A and
B)}. We adopted simple linear relation
\begin{equation}
F=c\cdot N,
\end{equation}
 where $N$ is source's total count, $F$ the flux in units
$10^{-15}$ erg cm$^{-2}$ s$^{-1}$, and $c=(4.97 \pm 0.05) \times
10^{-5}$ is calibration coefficient obtained from the fit.

We define the fractional error
\begin{equation}
f=\Bigg|\frac{F_{\mathrm{M\& H}}-F}{F_{\mathrm{M\& H}}}\Bigg|
\end{equation}
where $F_{\mathrm{M\&H}}$ is the H$\alpha$ flux from Miller \&
Hodge (1994), whereas $F$ is our measurement,  in order to get an
estimate of the accuracy of the obtained values. We find
$f_{\mathrm{max}}=0.50$ i $\bar{f}=0.16$.

Estimated fluxes for all sources (HII regions) and approximate
positions of source centers are given in Table 5.

\section{4. CONCLUSIONS}

We presented observations of Holmberg IX galaxy in narrow band
[SII] and H$\alpha$ filters. Our search for objects with an
enhanced [SII] emission -- possible supernova remnant candidates,
has yielded no sources of this class, besides M\&H 10-11 or HoIX
X-1. Nevertheless we identified 21 objects with significant
H$\alpha$ emission. Eleven sources (1--11) were previously
identified by Miller \& Hodge (1994) and one by Boone et al.
(2005) (source 12). Thus, we found nine new sources (13--21) --
uncatalogued HII regions. We estimated their H$\alpha$ fluxes and
gave their approximate positions.


\acknowledgements{This research has been supported by the Ministry
of Science and Technological Development of the Republic of Serbia
through projects:  No. 146002 "Astrophysical Spectroscopy of
Extragalactic Objects", No. 146003 "Physics of Sun and Stars",
 No. 146012 "Gaseous and Stellar Components of Galaxies: Interaction and
Evolution", project "The Rozhen Astronomical Observatory - Major
Facility for the South East European Region" funded by
UNESCO-ROSTE, and T\" ubitak project No. 09ARTT150-458-0 "Optical
Search for Supernova Remnants in M81, M82, NGC3077".}


\references

Arbutina, B., Uro\v{s}evi\'{c}, D. and Vukoti\'{c}, B.: 2007,
\journal{IAU Symp.}, \vol{237}, 391.

Blair, W.P. and Long, K.S.: 2004, \journal{Astrophys. J. Suppl.
Series},  \vol{155}, 101.

Boone, F., Brouillet, N., Huttemeister, S.,  Henkel, C., Braine,
 J., Bomans, D.J., Herpin, F., Banhidi, Z. and Albrecht, M.: 2005, \journal{Astron. Astrophys.},
\vol{429}, 129.

Gris\'{e}, F., Pakull, M. W. and Motch, C.: 2006, \journal{IAU
Symp.}, \vol{230}, 302.

Huang, Z. P., Thuan, T. X., Chevalier, R. A., Condon, J. J. and
Yin, Q. F.: 1994, \journal{Astrophys. J.}, \vol{424}, 114.

Karachentsev, I. D., Karachentseva, V. E., Huchtmeier, W. K.,
Makarov, D. I.: 2004, \journal{Astron. J.}, \vol{127}, 2031.

Karachentsev, I. D.  and  Kashibadze, O. G.: 2006,
\journal{Astrophysics}, \vol{49}, 3.

Matonick,  D. M. and Fesen, R. A.: 1997, \journal{Astrophys. J.
Suppl. Series}, \vol{112}, 49.

Miller, B.W.: 1995, \journal{Astrophys. J.}, \vol{446}, L75.

Miller, B.W. and Hodge, P.: 1994, \journal{Astrophys. J.},
\vol{427}, 656.

Monet, D. et al.: 1998, USNO-A2.0 - A catalog of astrometric
standards, U.S. Naval Observatory
({\texttt{http://tdc-www.harvard.edu/catalogs/ ua2.html}})

Sabbi, E. Gallagher, J.S., Smith, J.L., de Mello, D.F. and
Mountain, M.: 2008, \journal{Astrophys. J.}, \vol{676}, L113.

Rosa-Gonzales, D.: 2005,  \journal{Mon. Not. R. Astron. Soc.},
\vol{364}, 1304.

Yun, M.S., Ho, P.T.P. and Lo, K.Y.: 1994, \journal{Nature},
\vol{372}, 530.

\endreferences

}
\end{multicols}

\vfill\eject

{\ }



\naslov{OPTIQKA POSMATRANJA GRUPE GALAKSIJA M81 U USKIM FILTERIMA
[$\mathbf{SII}$] i $\mathbf{H}\alpha$: HOLMBERG $\mathbf{IX}$}


\authors{B. Arbutina$^{1}$, D. Ili\'{c}$^{1}$, K. Stavrev$^{2}$, D. Uro{\v s}evi{\' c}$^{1}$ B. Vukoti\'{c}$^3$ and D. Oni\' c$^1$}

\vskip3mm


\address{$^1$Department of Astronomy, Faculty of Mathematics,
University of Belgrade,\break Studentski trg 16, 11000 Belgrade,
Serbia}

\Email{arbo}{matf.bg.ac.rs}

\address{$^2$Institute of Astronomy, Bulgarian Academy of Sciences, 72 Tsarigradsko Shosse Blvd., \break BG-1784 Sofia, Bulgaria}

\address{$^3$Astronomical Observatory, Volgina 7, 11160 Belgrade,
Serbia}

\vskip.7cm


\centerline{UDK \udc}


\centerline{\rit Prethodno saopxtenje}

\vskip.7cm

\begin{multicols}{2}
{


\rrm U radu su predstavljena posmatranja oblizhnje patuljaste
galaksije Holmberg  $\mathrm{IX}$ u grupi galaksija vezanih za
M81. Posmatranja su izvrshena u martu i septembru 2008. godine
dvometarskim $\mathrm{RCC}$ teleskopom na NAO Rozhen, Bugarska,
korish\cj{}enjem uskih filtera $\mathrm{[SII]}$ i
$\mathrm{H}\alpha$.  Potraga za objektima sa pojachanom emisijom
$\mathrm{[SII]}$ u odnosu na $\mathrm{H}\alpha$ emisiju --
potencijalnim kandidatima za ostatke supernovih, nije rezultirala
novim objektima, pored  $\mathrm{M\&H}$ 10-11 ili $\mathrm{HoIX\
X}$-1, ali je zato detektovan jedan broj objekata sa znachajnom
$\mathrm{H}\alpha$ emisijom koji verovatno predstavljaju do sada
nepoznate $\mathrm{HII}$ regione.

}
\end{multicols}

\end{document}